\documentclass[aps,pra,showpacs]{revtex4}
\usepackage{graphicx,dcolumn,bm,xcolor,amsmath}

\begin{document}

\title{Propagation of three-dimensional bipolar ultrashort electromagnetic
pulses in an inhomogeneous array of carbon nanotubes}
\author{Eduard G. Fedorov}
\affiliation{Department of Biology, Technion-Israel Institute of Technology, Haifa 32000,
Israel}
\affiliation{Vavilov State Optical Institute, 199053 Saint Petersburg, Russia}
\author{Alexander V. Zhukov}
\affiliation{Singapore University of Technology and Design, 8 Somapah Road, 487372
Singapore}
\affiliation{Entropique Group Ltd., 3 Spylaw Street, Maori Hill, 9010 Dunedin, New Zealand}
\author{Roland Bouffanais}
\affiliation{Singapore University of Technology and Design, 8 Somapah Road, 487372
Singapore}
\author{Alexander P. Timashkov}
\affiliation{Saint Petersburg National Research University of Information Technologies,
Mechanics and Optics (ITMO University), 197101 Saint Petersburg, Russia}
\author{Boris A. Malomed}
\affiliation{Department of Physical Electronics, School of Electrical Engineering,
Faculty of Engineering, Tel Aviv University, 69978 Tel Aviv, Israel}
\affiliation{Saint Petersburg National Research University of Information Technologies,
Mechanics and Optics (ITMO University), 197101 Saint Petersburg, Russia}
\author{Herv{\'e} Leblond}
\affiliation{LUNAM Universit{\'e}, Universit{\'e} d'Angers, Laboratoire de Photonique
d'Angers, EA 4464, 2 Boulevard Lavoisier, 49000 Angers, France}
\author{Dumitru Mihalache}
\affiliation{Academy of Romanian Scientists, 54 Splaiul Independentei, Bucharest,
RO-050094, Romania}
\affiliation{Horia Hulubei National Institute of Physics and Nuclear Engineering,
Magurele, RO-077125, Romania}
\author{Nikolay N. Rosanov}
\affiliation{Vavilov State Optical Institute, 199053 Saint Petersburg, Russia}
\affiliation{Saint Petersburg National Research University of Information Technologies,
Mechanics and Optics (ITMO University), 197101 Saint Petersburg, Russia}
\affiliation{Ioffe Physical-Technical Institute, Russian Academy of Sciences, 194021
Saint Petersburg, Russia}
\author{Mikhail B. Belonenko}
\affiliation{Laboratory of Nanotechnology, Volgograd Institute of Business, 400048
Volgograd, Russia}
\affiliation{Volgograd State University, 400062 Volgograd, Russia}
\affiliation{Entropique Group Ltd., 3 Spylaw Street, Maori Hill, 9010 Dunedin, New Zealand}
\date{\today }

\begin{abstract}
We study the propagation of three-dimensional (3D) bipolar ultrashort
electromagnetic pulses in an inhomogeneous array of semiconductor carbon
nanotubes. The heterogeneity is represented by a planar region with an
increased concentration of conduction electrons. The evolution of the
electromagnetic field and electron concentration in the sample are governed
by the Maxwell's equations and continuity equation. In particular,
non-uniformity of the electromagnetic field along the axis of the nanotubes
is taken into account. We demonstrate that, depending on values of
parameters of the electromagnetic pulse approaching the region with the
higher electron concentration, the pulse is reflected from the region or
passes it. Specifically, our simulations demonstrate that, after interacting
with the higher-concentration area, the pulse can propagate steadily,
without significant spreading. The possibility of such ultrashort
electromagnetic pulses propagating in arrays of carbon nanotubes over
distances significantly exceeding characteristic dimensions of the pulses
makes it possible to consider them as 3D solitons.
\end{abstract}

\maketitle


\section{Introduction}


Nowadays, carbon nanotubes (CNTs)---quasi-one-dimensional macromolecules of
carbon---are considered as promising objects with a potential for
applications to the development of the elemental base of modern electronics,
including nanocircuits usable in neurocomputers~\cite{1}. Strong interest in
these materials, starting from the moment of their discovery~\cite{2,3}, is
due to their unique physical properties (e.g., see Refs.~\cite{4,5,6,7,8,9}%
), which, in addition to the above-mentioned potential for the use in
electronics, pave the way to a wide range of possibilities for the creation
of ultra-strong composite materials, fuel cells, chemical sensors, and
optical devices (such as displays, LEDs, and transparent conductive
surfaces), etc. From the viewpoint of optoelectronic applications, specific
features of the electronic structure of CNTs are of unique interest. The
nonparabolicity of the dispersion law of conduction electrons (i.e., the
energy dependence on the quasimomentum) in nanotubes makes it possible to
observe a number of unique electromagnetic phenomena, including nonlinear
diffraction and self-focusing of laser beams~\cite{10,11}, as well as the
propagation of solitary electromagnetic waves~\cite{12} at field strengths
starting from $\sim 10^{3}-10^{4}$ V/cm. In this connection, it is relevant
to mention that possibilities offered by modern laser technologies for the
generation of powerful electromagnetic radiation with specified
properties---including ultrashort laser pulses with the duration on the
order of several half-cycles of field oscillations~\cite{13,14}---have
stimulated studies of the propagation of electromagnetic waves in various
novel media~\cite{15,16,17,18,19,20,21,22,23,23a,23b}, CNTs being one of
them.

The possibility of propagation of solitary electromagnetic waves in arrays
of CNTs has been first theoretically established in a one-dimensional (1D)
model based on an assumption of uniformity of the field along the nanotube
axis~\cite{12}. Actually, this approximation is only valid in a very narrow
range of values of the underlying parameters. Subsequent studies aimed at
investigating increasingly more realistic models describing the evolution of
the electromagnetic field in nanotube arrays, taking into regard various
physical factors affecting the dynamics of the electromagnetic wave. Adding
the complexity to the model's framework proceeded in several directions: i)
increasing the dimensionality of the setting, ii) taking into account
non-uniformity of the electromagnetic wave field, and iii) including various
inhomogeneities of the medium. The investigation of these factors was
carried out both independently, to clarify the role of each factor, and,
subsequently, considering combined effects of multiple factors.

Naturally, the first factor considered to make the model more realistic was
the dimensionality of the model. The propagation of electromagnetic waves in
arrays of CNTs was studied in the framework of 1D~\cite{12,24}, 2D~\cite%
{25,26,27,28}, and 3D~\cite{29,30} models. It was established that bipolar
ultrashort electromagnetic pulses propagate, in a stable fashion, in the
form of \textquotedblleft breather-like" light bullets over distances far
exceeding the characteristic size of the pulses along the direction of their
motion. The topicality of studying the propagation of electromagnetic waves
in arrays of nanotubes in the inhomogeneous model is due, in particular, to
the need to take transverse diffraction into account. Moreover, as in
reality, a solitary electromagnetic wave in a real physical system is
bounded in all directions, which implies non-uniformity of the field in any
direction. Thus, the necessity of constructing a 3D model for the evolution
of the electromagnetic field is obvious. In particular, non-uniformity of
the field of the electromagnetic pulse along the axis of the nanotubes is an
inherent ingredient of the 3D model. The construction of a model involving
the latter feature was first reported in Refs.~\cite{28,29,30}. In
particular, the propagation of ultrashort pulses in homogeneous arrays of
CNTs has been investigated in 2D~\cite{28} and 3D~\cite{29} models. One of
the main results of these studies is the prediction of redistribution of
conduction electrons, leading to specific variations of the density of
conduction electrons. Among possible practical applications of this
phenomenon in micro- and nano-electronics, one can envisage the
manufacturing of highly accurate chemical sensors~\cite{29} based on
specifically designed arrays of semiconductor CNTs. As shown in Ref.~\cite%
{30}, such dynamical inhomogeneities in the electron subsystem of nanotubes
can also underlie a complex medium-mediated mechanism of interaction of
colliding electromagnetic pulses in the array of nanotubes.

Among factors that significantly affect the evolution of electromagnetic
waves in arrays of semiconductor CNTs, a noteworthy one is the presence of
various impurities and inhomogeneities. In Ref.~\cite{31}, the propagation
of a bipolar electromagnetic pulse in the 2D geometry, and in the presence
of a multilevel impurity uniformly distributed over the sample, was
investigated. It was shown that doping the medium with an impurity of this
type leads to a modification of the characteristics of the propagating pulse
as compared to the case of propagation in a \textquotedblleft pure" sample.

Specific sample-doping format deserve particular consideration---in
particular, localized introduction of impurities, limited to a certain part
of the sample's volume, which implies the creation of an inhomogeneity. Such
defects of the medium, which are not initially associated with the action of
the electromagnetic field, can be, for example, implemented as metallic
inclusions or layers containing an increased concentration of conduction
electrons relative to the concentration in the homogeneous part of the
sample. This kind of heterogeneity can emerge either as a consequence of
technological failures, at the stage of sample manufacturing, or as a result
of purposeful formation of the inhomogeneity with intended properties. The
features of the interaction of 2D unipolar \textquotedblleft light bullets"
with metallic inclusions in nanotube arrays have been theoretically studied
in Refs.~\cite{32,33}. The interaction of a bipolar ultrashort
electromagnetic pulse with a layer of an increased electron concentration in
the 2D model was studied both under the assumption of the uniformity of the
field along the nanotube axis~\cite{34}, and also taking into account
non-uniformity of the field~\cite{35}. As a result, the selective nature of
the interaction of the pulse with the static inhomogeneity of the medium has
been established: a decrease of the pulse's duration and increase of its
amplitude facilitates its passage through a layer of an increased electron
concentration, while pulses with a longer duration and smaller amplitude can
be reflected by this layer.

Thus, the propagation of electromagnetic pulses in inhomogeneous
media---taking into account the non-uniformity of the pulses' field---are of
significant interest. Indeed, there is a need to search for physical effects
that can be used for the development of new components of the elemental base
of optoelectronics, developing schemes for optical information processing
and nondestructive testing systems, etc. Some peculiarities of light-matter
interaction mentioned above may be employed for the development of
ultra-fast optical transistors, switches, logic elements, transmission and
signal delay lines, soliton memory elements \cite{35a}, using ultra-short
pulses as bits of data. Different outcomes of data processing operations may
be associated with different regimes of the pulse propagation in the medium
(e.g., different outcomes of the collision of a pulse with a layer of
increased conductivity, embedded in the bulk of a semiconductor structure).
It is therefore topical to implement such concepts using some of the most
promising available materials, such as, in particular, graphene-based
materials. In this connection, it is relevant to address effects of the
static inhomogeneity in the electron subsystem of the array of semiconductor
CNTs on the propagation of an ultrashort bipolar electromagnetic pulses in
the most realistic 3D geometry. In this work, we consistently take into
account factors that affect the dynamics of the ultrashort laser pulses,
generalizing previously considered particular cases in the framework of an
integrated model.


\section{The system's configuration and key assumptions}

\begin{figure}[tbp]
\includegraphics[width=0.6\textwidth]{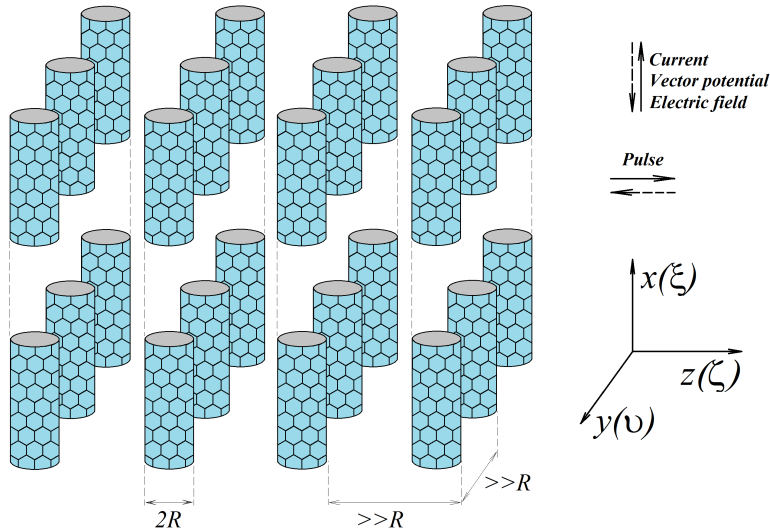}
\caption{The schematic plot of the setup and the associated coordinate
system.}
\label{fig1}
\end{figure}

We consider the propagation of a solitary electromagnetic wave in a
volumetric array of a single-walled semiconductor carbon nanotubes (CNTs)
embedded in a homogeneous dielectric medium. The nanotubes considered here
are of the \textquotedblleft zigzag" type $(m,0)$, where integer $m$ (not a
multiple of three for semiconductor nanotubes) determines their radius, $R=mb%
\sqrt{3}/2\pi $, with $b=1.42\times 10^{-8}\mathrm{~cm}$ being the distance
between neighboring carbon atoms~\cite{4,5,6,7}. The CNTs are arranged in
such a way that their axes are parallel to the common $x$-axis, and
distances between adjacent nanotubes are much larger than their diameter.
The latter assumption allows one to neglect the interaction between CNTs~%
\cite{36}. Moreover, it allows us to consider the system as an electrically
quasi-1D one, in which electron tunneling between neighboring nanotubes may
be neglected, and electrical conductivity is possible only along the axis of
the nanotubes. We define the configuration of the system in such a way that
the pulse propagates through the CNT array in the direction perpendicular to
their axes (for the definiteness' sake, along the $z$-axis), while the
electric component of the wave field, $\mathbf{E}=\{E,0,0\}$, is collinear
with the $x$-axis (see Fig.~\ref{fig1}).

For a wide range of values of the system's parameters, the characteristic
distance at which an appreciable change in the field of a bipolar
electromagnetic pulse occurs, is significantly greater than both the
distance between neighboring nanotubes and the length of the
conduction-electron's path along the axis of the nanotubes. On the other
hand, with the nanotube radius $R\approx 5.5\times 10^{-8}\mathrm{~cm}$ and $%
m=7$, the characteristic distance between the nanotubes, sufficient to
exclude the overlap of the electron wave functions of adjacent ones, i.e.,
substantially exceeding $R$, may still be negligibly small in comparison
with the wavelength $\lambda $ of the electromagnetic radiation. In
particular, this assumption is definitely valid for the infrared radiation,
with $\lambda >1$ $\mathrm{\mu }$m. Under this condition, the nanotube
array, which is a discrete structure at the microscopic level, may
nevertheless be considered as a quasi-uniform medium for the propagation of
electromagnetic waves. In this case, for length scales comparable to the
pulse's dimensions, the array of CNTs may be considered as a uniform
continuous medium. In other words, the electromagnetic field in the
system---specifically, ultra-short pulses carried by the infrared
wavelength---are not affected by the discrete structure in the medium, there
being no scattering (or even recurrent scattering) of electromagnetic
radiation on inhomogeneities or irregularities of the CNT volumetric array.
Of course, the scattering will appear in the framework of a microscopic
theory (see, e.g., Ref. \cite{Lagendijk}), which should be a subject for a
separate work.

Another assumption that we adopt here concerns the time duration of the
electromagnetic pulse, $T_{S}$, the relaxation time of the conduction
current along the nanotube axis, $t_{\mathrm{rel}}$, and also the time of
the observation of the light propagation in the system, $t$ ---we assume $%
T_{S}\ll t<t_{\mathrm{rel}}$. If this condition is met, it is possible to
neglect field decay, thus enabling the collisionless approximation in
describing the lossless evolution of the field~\cite{12}.


\section{Governing equations}

Given the orientation of the coordinate system relative to the nanotube
axis, as defined in Fig.~\ref{fig1}, the electron energy spectrum for CNTs
takes the form~\cite{29, 30}
\begin{equation}
\epsilon (p_{x},s)=\gamma _{0}\sqrt{1+4\cos \left( p_{x}\frac{d_{x}}{\hbar }%
\right) \cos \left( \pi \frac{s}{m}\right) +4\cos ^{2}\left( \pi \frac{s}{m}%
\right) },  \label{1}
\end{equation}%
where the electron quasimomentum is $\mathbf{p}=\left\{ p_{x},s\right\} $, $%
s=1,2,\dots ,m$ being an integer characterizing the momentum quantization
along the perimeter of the nanotube, with $m$ being the number of hexagonal
carbon cycles which form the circumference of the CNT, $\gamma _{0}$ is the
overlap integral, and $d_{x}=3b/2$.

\subsection{Equation for the vector potential}

The electromagnetic field in the CNT\ array is governed by Maxwell's
equations~\cite{37,38}, from which, taking into account the Lorentz gauge
condition, we obtain the wave equation for the spatiotemporal evolution of
the vector field potential:
\begin{equation}
\frac{\varepsilon }{c^{2}}\frac{\partial ^{2}\mathbf{A}}{\partial t^{2}}-
\frac{\partial ^{2}\mathbf{A}}{\partial x^{2}}-\frac{\partial ^{2}\mathbf{A}
}{\partial y^{2}}-\frac{\partial ^{2}\mathbf{A}}{\partial z^{2}}=\frac{4\pi
}{c}\mathbf{j},  \label{2}
\end{equation}%
with $\mathbf{A}=\left\{ A,0,0\right\} $, where $\mathbf{j}=\left\{
j,0,0\right\} $ is the current density, $c$ the speed of light in vacuum,
and $\varepsilon $ the average relative dielectric constant of the medium
(see, e.g., Refs.~\cite{25,27,28,29,30}).

We emphasize that the system under consideration has a nonzero electric
conductivity only along the $x$--axis, while in the $\left( y,z\right) $
plane the current is absent due to the negligible coupling between
neighboring nanotubes. Thus, since the second and third components of the
conduction current $\mathbf{j}$ are zero, Eq.~\eqref{2} admits the existence
of zero solutions for the second and third components of the vector
potential. We use this fact to define the vector potential as being
collinear to axes of the nanotubes.

The conduction current density $j$ along the nanotube axis is determined by
applying the approach used in Refs.~\cite{39, 40}, which yields%
\begin{equation}
j=2e\sum_{s=1}^{m}\int\limits_{-\pi \hbar /d}^{\pi \hbar
/d}v_{x}f(p_{x},s)dp_{x},  \label{3}
\end{equation}%
where $e<0$ is the electron charge, $v_{x}$ is the electron velocity and $%
f(p_{x},s)$ is the electron distribution function with respect to
quasimomenta $p_{x}$ and numbers $s$ characterizing the quantization of the
electron's momentum along the perimeter of a nanotube. The integration over
the quasimomentum in Eq. (\ref{3}) is carried out within the first Brillouin
zone.

Using the expression for the electron energy~\eqref{1} to determine their
velocity as $v_{x}=\partial \epsilon (p_{x},s)/\partial p_{x}$, and taking
into account the electron distribution $f(p_{x},s)$ according to the
Fermi--Dirac statistics, we derive from Eq.~\eqref{3}, an expression for the
current density (for more details see Ref.~\cite{30}):
\begin{equation}
j=-en\frac{d_{x}}{\hbar }\gamma _{0}\sum_{r=1}^{\infty }G_{r}\sin \left[ r%
\frac{d_{x}}{\hbar }\left( A\frac{e}{c}+e\int\limits_{0}^{t}\frac{\partial
\phi }{\partial x}dt^{\prime }\right) \right] ,  \label{4}
\end{equation}%
where $n=n(x,y,z,t)$ is the local value of the concentration of conduction
electrons, $\phi $ is the scalar potential (self-consistent equations for
the quantities $n$ and $\phi $ are derived in Sec.~III.B and Sec.~III.C,
respectively), and coefficients $G_{r}$ are given by
\begin{equation}
G_{r}=-r\frac{\displaystyle\sum\limits_{s=1}^{m}{\displaystyle\frac{\delta
_{r,s}}{\gamma _{0}}}\displaystyle\int_{-\pi }^{+\pi }\cos (r\kappa )\left\{
1+\exp \left[ \frac{\theta _{0,s}}{2}+\sum_{q=1}^{\infty }\theta _{q,s}\cos
\left( q\kappa \right) \right] \right\} ^{-1}\text{d}\kappa }{\displaystyle%
\sum_{s=1}^{m}\int_{-\pi }^{+\pi }\left\{ 1+\exp \left[ \frac{\theta _{0,s}}{%
2}+\sum_{q=1}^{\infty }\theta _{q,s}\cos \left( q\kappa \right) \right]
\right\} ^{-1}\text{d}\kappa }.  \label{5}
\end{equation}%
Here $\theta _{r,s}=\delta _{r,s}(k_{B}T)^{-1}$, while $T$ is the
temperature, $k_{B}$ the Boltzmann constant, and $\delta _{r,s}$ are
coefficients of the Fourier decomposition~\cite{41} of spectrum~\eqref{1}:
\begin{equation}
\delta _{r,s}=\frac{d_{x}}{\pi \hbar }\int_{-\pi \hbar /d_{x}}^{-\pi \hbar
/d_{x}}\epsilon (p_{x},s)\cos \left( r\frac{d_{x}}{\hbar }p_{x}\right) \text{
d}p_{x}.  \label{6}
\end{equation}

The evolution of the vector potential of the field in the system is
determined by the projection of Eq.~\eqref{2} onto the nanotube axis, which,
taking into account expression~\eqref{4} and after introducing dimensionless
variables, takes the following form:
\begin{equation}
\frac{\partial ^{2}\Psi }{\partial \tau ^{2}}-\left( \frac{\partial ^{2}\Psi
}{\partial \xi ^{2}}+\frac{\partial ^{2}\Psi }{\partial \upsilon ^{2}}+\frac{
\partial ^{2}\Psi }{\partial \zeta ^{2}}\right) +\eta \sum_{r=1}^{\infty
}G_{r}\sin \left[ r\left( \Psi +\int\limits_{0}^{\tau }\frac{\partial \Phi }{
\partial \xi }d\tau ^{\prime }\right) \right] =0,  \label{7}
\end{equation}%
where $\eta =n/n_{\mathrm{bias}}=\eta (\xi ,\upsilon ,\zeta ,\tau )$ is the
reduced (dimensionless) density of conduction electron, $n_{\mathrm{bias}}$
is the concentration of conduction electrons in the homogeneous part of the
sample in the absence of electromagnetic fields, $\Psi =Aed_{x}/\left(
c\hbar \right) $ is the projection of the scaled vector potential onto the $%
x $-axis, $\Phi =\phi \sqrt{\varepsilon }ed_{x}/(c\hbar )$ is the
dimensionless scalar potential, $\tau =\omega _{0}t/\sqrt{\varepsilon }$ is
the scaled time, $\xi =x\omega _{0}/c$, $\upsilon =y\omega _{0}/c$ and $%
\zeta =z\omega _{0}/c$ are the scaled coordinates, and
\begin{equation}
\omega _{0}\equiv 2\frac{|e|d_{x}}{\hbar }\sqrt{\pi \gamma _{0}n_{\mathrm{\
bias}}}.  \label{8}
\end{equation}%
Thus, Eq.~\eqref{7} describes the evolution of the vector potential of the
self-consistent electromagnetic field in the CNT array: the field is fully
determined by the density of the conduction current [see Eq.~\eqref{2}], and
the conduction current is, in turn, affected by the field [see Eq.~\eqref{4}%
].

\subsection{Equation for the electron density}

In the general case, the electromagnetic field in the system under
consideration is non-uniform in space. Indeed, the field of an ultrashort
electromagnetic pulse propagating in an array of nanotubes is localized at
each moment in a small (moving) region of space. The non-uniformity is
invisible on the scale of the nanotube radius $\sim 5\times 10^{-8}\mathrm{\
~cm}$, or even for the distance between neighboring nanotubes $\sim
10^{-7}-10^{-6}\mathrm{~cm}$. However, it is significant at the wavelength
scales of the infrared radiation, $\lambda $ $\sim 1\mathrm{\ \mu }$m.

The spatial non-uniformity of the field along the nanotube axis determines
the dependence of the current density on coordinate $x$, as it follows from
expression~\eqref{4} for the current density. Since the total charge in the
sample is conserved, and the change in its bulk density $\rho =en$ obeys the
continuity equation $\nabla \mathbf{j}+\partial \rho /\partial t=0$~\cite%
{37,38}, the non-uniformity of the current density causes a temporal change
in electron density, as per
\begin{equation}
\frac{\partial n}{\partial t}=-\frac{1}{e}\frac{\partial j}{\partial x}.
\label{9}
\end{equation}%
We stress that, as the system considered here is supposed to be electrically
quasi-one-dimensional, i.e., the conductivity is only effective along the
nanotubes axis, given the negligible overlap of the electron wave functions
of neighboring nanotubes, the field non-uniformity along directions
orthogonal to the nanotube axes does not affect the distribution of the
electron concentration in the sample.

Substituting Eq.~\eqref{4} for the projection of the current density onto
the axis of nanotubes into Eq.~\eqref{9}, and passing to the dimensionless
notation (the same as in Ref.~\cite{30}), we obtain an equation governing
the evolution of the electron concentration under the action of the
electromagnetic pulse:
\begin{equation}
\frac{\partial \eta }{\partial \tau }=\alpha \sum_{r=1}^{\infty }G_{r}\frac{
\partial }{\partial \xi }\left\{ \eta \sin \left[ r\left( \Psi
+\int\limits_{0}^{\tau }\frac{\partial \Phi }{\partial \xi }d\tau ^{\prime
}\right) \right] \right\} ,  \label{10}
\end{equation}%
with $\alpha \equiv d_{x}\gamma _{0}\sqrt{\varepsilon }/c\hbar $, the other
quantities being defined in Eq.~\eqref{7}.

\subsection{Equation for the scalar potential field}


The system as a whole being electro-neutral, the redistribution of the
electron concentration in the sample, due to the action of the non-uniform
field along the axis of the nanotubes, is equivalent to appearance of
regions of high and low electron concentration relative to the initial
equilibrium distribution, $n_{0}=n(\xi ,\upsilon ,\zeta ,\tau _{0})$, taken
(at the the initial time $\tau _{0}$) prior to the entrance of the an
electromagnetic pulse into the sample. Thus, the local concentration of
electrons, $n(\xi ,\upsilon ,\zeta ,\tau )$, may be represented as the sum
of the initial equilibrium value $n_{0}$ and the concentration of the
\textquotedblleft additional" charge, $\delta n(\xi ,\upsilon ,\zeta ,\tau
)=n-n_{0}$, with density $\delta \rho =e\delta n=e(n-n_{0})$. Note that $%
\delta \rho \neq 0$ implies a local imbalance between the negative charge of
free electrons and the positive charge of holes. The local imbalanced charge
perturbs the distribution of the field according to the driven wave equation
for the scalar potential, which follows from Maxwell's equations~\cite{37,38}%
:
\begin{equation}
\frac{\varepsilon }{c^{2}}\frac{\partial ^{2}\phi }{\partial t^{2}}-\left(
\frac{\partial ^{2}\phi }{\partial x^{2}}+\frac{\partial ^{2}\phi }{\partial
y^{2}}+\frac{\partial ^{2}\phi }{\partial z^{2}}\right) =\frac{4\pi }{
\varepsilon }\delta \rho .  \label{11}
\end{equation}%
Using the same dimensionless notations as above, Eq.~\eqref{11} can be
written as
\begin{equation}
\frac{\partial ^{2}\Phi }{\partial \tau ^{2}}-\left( \frac{\partial ^{2}\Phi
}{\partial \xi ^{2}}+\frac{\partial ^{2}\Phi }{\partial \upsilon ^{2}}+\frac{
\partial ^{2}\Phi }{\partial \zeta ^{2}}\right) =\beta (\eta -\eta _{0}),
\label{12}
\end{equation}%
where $\beta =1/\alpha =c\hbar /\left( d_{x}\gamma _{0}\sqrt{\varepsilon }%
\right) $ [see Eq.~\eqref{10}], and $\eta _{0}=n_{0}/n_{\mathrm{bias}}=\eta
(\xi ,\upsilon ,\zeta ,\tau _{0})$ is the dimensionless local value of the
concentration of conduction electrons at the initial instant of time in the
absence of the field.

Thus, the evolution of the field in the CNT\ array, taking into account the
redistribution of the conduction-electron density, is governed by the system
of equations \eqref{7}, \eqref{10}, and \eqref{12}, which provide a
self-consistent model for the evolution of the electromagnetic field and
electronic subsystem in the array.

\subsection{The localization of the electromagnetic pulse}


Upon obtaining the numerical solution to equations \eqref{7}, \eqref{10},
and \eqref{12} it is possible to calculate the electric field, as $\mathbf{E}%
=-c^{-1}\partial \mathbf{A}/\partial t-\nabla \phi $ (see, e.g., Refs.~\cite%
{37, 38}). Taking into account that the vector potential has a nonzero
component only along the nanotube axis (see the description of the system
configuration above), one can write expressions for the components of the
electric field as follows:

\begin{equation}
E_{x}=E_{0}\left( \frac{\partial \Psi }{\partial \tau }+\frac{\partial \Phi
}{\partial \xi }\right) ,\quad E_{y}=E_{0}\frac{\partial \Phi }{\partial
\upsilon },\quad E_{z}=E_{0}\frac{\partial \Phi }{\partial \zeta },
\label{13}
\end{equation}%
where $E_{0}\equiv -\hbar \omega _{0}/ed_{x}\sqrt{\varepsilon }$. Thus, the
electric field in the CNT array is not, generally, collinear with the
nanotube axis. However, the $\ y$ and $z$ components of the electric field,
orthogonal to the nanotube axis, do not affect the dynamics of electrons,
due to the absence of conductivity of the system in these directions. Thus,
only component $E_{x}$, which affects the dynamics of the electronic
subsystem, is relevant to the description of the solitary electromagnetic
wave. The energy density of this component is%
\begin{equation}
I=E_{x}^{2}=I_{0}\left( \frac{\partial \Psi }{\partial \tau }+\frac{\partial
\Phi }{\partial \xi }\right) ^{2},  \label{14}
\end{equation}%
where $I_{0}=E_{0}^{2}$, and we have made use of the first expression in Eq.
(\ref{13}). The position of a local maxima of this quantity identifies the
instantaneous location of the ultrashort electromagnetic pulse.


\section{Non-uniformity of the electron concentration}

We assume that the CNT array contains a localized inhomogeneity, in the form
of a region with an increased concentration of conduction electrons,
exceeding the bulk concentration $n_{\mathrm{bias}}$ in the homogeneous
sample. As mentioned above, such a local defect can be created by
introducing donor impurities at the stage of the fabrication of the sample,
subject to the condition of the electro-neutrality of the entire system.
Accordingly, the initial electron density (in the absence of an
electromagnetic pulse) is $n_{0}=n(\xi ,\upsilon ,\zeta ,\tau _{0})$. We
stress that, in the absence of the electromagnetic pulse, each segment of
the sample is locally electro-neutral, even if the electron concentration is
inhomogeneous. Namely, in the region of increased electron density, the hole
concentration is higher too, compensating the charge of the free electrons.

\begin{figure}[tbp]
\includegraphics[width=0.5\textwidth]{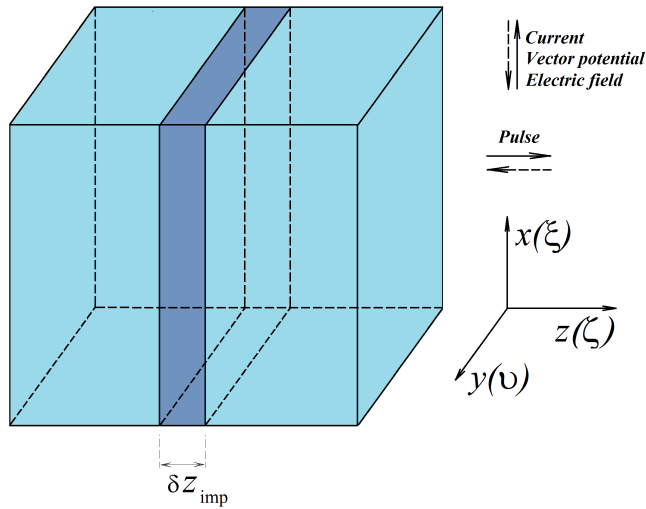}
\caption{The layer of increased concentration of conduction electrons.}
\label{fig2}
\end{figure}

We assume that the region of the increased electron concentration is a
narrow layer parallel to the nanotube axis, and its thickness $\delta z_{
\mathrm{imp}}$ is much smaller than the spatial dimension of the
electromagnetic pulse in the direction of its propagation along the $z$%
-axis. In addition, we also consider this narrow layer as being indefinitely
extended in the $x$ and $y$, as shown in Fig.~\ref{fig2}.

The respective dimensionless electron concentration in the sample in the
absence of the electromagnetic field is approximated by the Gaussian profile
(see Ref.~\cite{35}):
\begin{equation}
\eta (\xi ,\upsilon ,\zeta ,\tau _{0})=\eta (\zeta )=1+\left( \eta _{\mathrm{%
\ imp}}^{\mathrm{max}}-1\right) \exp \left\{ -\left( \frac{\zeta }{\delta
\zeta _{\mathrm{imp}}}\right) ^{2}\right\} ,  \label{15}
\end{equation}%
where $\eta _{\mathrm{imp}}^{\mathrm{max}}=n_{\mathrm{imp}}^{\mathrm{max}%
}/n_{\mathrm{bias}}$, $n_{\mathrm{imp}}^{\mathrm{max}}$ being the maximum
electron concentration in the region of inhomogeneity, and $\delta \zeta _{
\mathrm{imp}}$ is a dimensionless parameter determined by the characteristic
half-thickness of the region of the increased electron concentration, $%
\delta \zeta _{\mathrm{imp}}=\omega _{0}\delta z_{\mathrm{imp}}/c$. The
concentration of conduction electrons is assumed to be constant in the $%
\left( x,y\right) $ plane.

The initial electron density in the system determines the corresponding
scalar potential. Taking into account the fact that $\eta \equiv \eta _{0}$
holds initially, the right-hand side of Eq.~\eqref{12} vanishes. As a
result, Eq.~\eqref{12} produces a constant solution. Since the scalar
potential is always determined up to an arbitrary constant~\cite{37,38}, its
initial value may be fixed to be zero:
\begin{equation}
\Phi (\xi ,\upsilon ,\zeta ,\tau _{0})=0,  \label{16}
\end{equation}%
which we assume to be the initial distribution of the scalar potential in
the system.


\section{The initial form of the electromagnetic pulse}

We now assume that the electromagnetic pulse propagates in the array of
CNTs, with the $\xi $-component of the dimensionless vector potential at the
initial instant of time $\tau =\tau _{0}$ defined as follows:
\begin{equation}
\Psi (\xi ,\upsilon ,\zeta ,\tau _{0})=\Psi _{B}(\zeta ,\tau _{0})\exp \left[
-\frac{(\xi -\xi _{0})^{2}+(\upsilon -\upsilon _{0})^{2}}{w_{0}^{2}}\right] ,
\label{17}
\end{equation}%
where $\Psi _{B}(\zeta ,\tau _{0})$ is the $\xi $-component at $\xi =\xi
_{0} $ and $\upsilon =\upsilon _{0}$, with $\xi _{0}$ and $\upsilon _{0}$
being dimensionless pulse's coordinates, along the $\xi $- and $\upsilon $%
-axes, respectively, at the initial instant of time, and the initial
transverse half-width $w_{0}$ of the pulse.

Profile $\Psi _{B}(\zeta ,\tau _{0})$ is chosen as a breather of the
sine-Gordon equation, i.e., a non-topological oscillating soliton~\cite{42}:
\begin{equation}
\Psi _{B}(\zeta ,\tau _{0})=4\arctan \left\{ \left( \frac{1}{\Omega ^{2}}
-1\right) ^{1/2}\frac{\sin \chi }{\cosh \mu }\right\} ,  \label{18}
\end{equation}%
where

\begin{align}
\chi & \equiv \sigma \Omega \frac{\tau _{0}-(\zeta -\zeta _{0})U}{\sqrt{
1-U^{2}}},  \label{19} \\
\mu & \equiv \sigma \left[ \tau _{0}U-(\zeta -\zeta _{0})\right] \sqrt{\frac{
1-\Omega ^{2}}{1-U^{2}}},  \label{20}
\end{align}%
$U=u/v_{0}$ is the ratio of the initial propagation velocity $u$ of the
electromagnetic pulse (breather) along the $\zeta $-axis within the
sine-Gordon approximation, and the linear speed of light in the medium, $%
v_{0}=c/\sqrt{\varepsilon }$. Further, $\zeta _{0}$ is the dimensionless
coordinate of the breather along the $\zeta $-axis at moment $\tau =\tau
_{0} $, $\Omega <1$ is a free parameter, which determines the breather's
oscillation frequency (scaled by frequency $\omega _{B}=\omega _{0}\Omega $
in physical units), and $\sigma =\sqrt{G_{1}}$ [coefficients $G_{j}$ are
calculated as per Eq.~\eqref{5}].

The basic argument in favor of the choice of the initial condition~in the
form of Eqs. \eqref{17} and~\eqref{18} is that the equation for the vector
potential~\eqref{7} may be considered as a non-uniform generalization of the
sine-Gordon equation. As the sine-Gordon equation gives rise to breather
solutions in the form~of Eq. \eqref{18}, it is reasonable to assume the
possibility of the propagation of solitary waves in a form close to
breathers. This assumption is amply justified by results reported in Refs.~%
\cite{25,26,27,28,29,30}. The second factor in Eq.~\eqref{17} corresponds to
the Gaussian distribution of the field in the plane $(\xi ,\zeta )$
perpendicular to the propagation direction of the electromagnetic pulse. The
choice of the Gaussian distribution for this field is justified by a wide
range of applicability of the Gaussian waveforms~\cite{43,44,45,46,47}.

The component of the electric field \eqref{13} along the nanotube axis,
taking into account the expression~\eqref{17} at the initial instant of
time, has the form of
\begin{eqnarray}
E_{x} &=&4E_{0}\frac{\sigma \sqrt{1-\Omega ^{2}}}{\sqrt{1-U}^{2}}\left\{
\frac{\cos \chi \cosh \mu -U\left( \Omega ^{-2}-1\right) ^{1/2}\sin \chi
\sinh \mu }{\cosh ^{2}\mu +\left( \Omega ^{-2}-1\right) \sin ^{2}\chi }
\right\}  \notag \\
&\times &\exp \left[ -\frac{(\xi -\xi _{0})^{2}+(\upsilon -\upsilon
_{0})^{2} }{w_{0}^{2}}\right] .  \label{21}
\end{eqnarray}%
Equations~\eqref{17}--\eqref{21} describe a short wave packet, consisting of
a carrier wave and an envelope. The carrier, which accounts for the
internal oscillations~\cite{25,29} of the pulse, is determined by the
oscillating behavior of the function $\sin \chi $, while the envelope
accounts for the exponential behavior of the function $\cosh \mu $. In the
case of a few-cycle pulse, variation scales of both the envelope and carrier
of the pulse have the same order of magnitude, hence 
its profile varies periodically with the frequency of the carrier.
The pulse given by Eq. \eqref{21} is categorized as a \textquotedblleft
bipolar" one, as the sign of this field component changes periodically.

We emphasize that the initial parameters of the electromagnetic pulse at $%
\tau =\tau _{0}$ are given under the assumption that the pulse is still
located at a sufficient distance from the inhomogeneity layer, where the
scaled concentration of conduction electrons~\eqref{15} is different from $1$%
. From the experimental viewpoint, the effective optical frequency, $\omega
_{\mathrm{opt}}$, and the characteristic duration of the pulse, $T_{S}$, are
relevant parameters characterizing the shape of the electromagnetic pulse~%
\cite{35}.

The optical frequency $\omega _{\mathrm{opt}}$ is determined as follows. As
said above, the internal vibrations of breather~\eqref{18} are
represented by the function $\sin \chi $. We represent the argument (see Eq.~%
\eqref{19}) of this function in the form
\begin{equation}
\chi =\omega _{c}t_{0}+k_{\mathrm{wave}}(z-z_{0}),  \label{22}
\end{equation}%
where $k_{\mathrm{wave}}$ is the wave vector. In dimensionless form and
using Eq.~\eqref{19}, we have:
\begin{equation}
\chi =\frac{\omega _{0}}{\sqrt{\varepsilon }}\frac{\sigma \Omega }{\sqrt{
1-U^{2}}}t_{0}-U\frac{\sigma \Omega }{\sqrt{1-U^{2}}}(z-z_{0}).  \label{23}
\end{equation}%
By comparing Eq.~\eqref{19} and Eq.~~\eqref{20}, we obtain the following
expression for the carrier-wave's frequency:
\begin{equation}
\omega _{c}=\frac{\omega _{0}}{\sqrt{\varepsilon }}\frac{\sigma \Omega }{
\sqrt{1-U^{2}}}.  \label{24}
\end{equation}

It must, however, be noticed that frequency $\omega _{c}$ coincides with the
frequency at which the optical spectrum reaches its maximum in the
slowly-varying-envelope approximation only. For few- and single-cycle
pulses, the latter frequency is larger than the former, the ratio between
them increasing as the number of cycles decreases, up to $\approx 1.66$.
Further, due to the strong nonlinear behavior, the spectrum is not conserved
in the course of the propagation, and the frequency at which it reaches its
maximum oscillates. According to Eqs.~\eqref{17}--\eqref{21}, i.e., in the
framework of the sine-Gordon approximation, the amplitude of the
oscillations can reach $\pm 0.14\omega _{c}$.

Duration $T_{S}$ of the electromagnetic pulse (wave packet) with vector
potential~\eqref{18} is determined by the factor $\cosh \mu $. The usual
FWHM definition of $T_{S}$ is the time during which the instantaneous
amplitude of the \textquotedblleft running" envelope, measured at a fixed
point, exceeds half of its peak value. In the few-cycle regime, the
definition of $T_{S}$ should be implemented numerically, using, e.g., the
standard deviation.
Further, even this standard definition can give rise to ambiguities, as
there is some discrepancy between the durations of fields $\Psi $ and $E_{x}$%
, the ratio of which may become $1.3$ in the single-cycle regime.

Therefore, it is more convenient do define the pulse duration in terms of
the slowly-varying-envelope approximation. According to the definition %
\eqref{20} of $\mu $, the role of the characteristic normalized duration of
the pulse may be played by%
\begin{equation}
\tau _{S}=\frac{1}{\sigma U}\sqrt{\frac{1-U^{2}}{1-\Omega ^{2}}}.
\label{tauS}
\end{equation}%
The fact that $\tau _{S}\sim 1/U$ at $U\rightarrow 0$ corresponds to the
slowly-varying-envelope approximation limit, the expression for the pulse in
this limit being
\begin{equation}
E_{xB}=E_{0}\frac{\partial \Psi _{B}}{\partial \tau }=4E_{0}\frac{\sigma
\sqrt{1-\Omega ^{2}}}{\sqrt{1-U^{2}}}\left\{ \frac{\cos \chi \cosh \mu }{
\cosh ^{2}\mu +\left( \Omega ^{-2}-1\right) \sin ^{2}\chi }\right\} ,
\label{ESVEA}
\end{equation}%
at $\xi =\xi _{0}$, $\zeta =\zeta _{0}$. It is close to the usual
sech-shaped pulse, which would be obtained by neglecting the term $\sin
^{2}\chi $ in the denominator:
\begin{equation}
E_{xB\mathrm{sech}}=4E_{0}\frac{\sigma \sqrt{1-\Omega ^{2}}}{\sqrt{1-U^{2}}}
\frac{\cos \chi }{\cosh \mu },  \label{Esech}
\end{equation}%
but does not coincide with it. Furthermore, an explicit result for the
envelope does not follow straightforwardly from Eq.~\eqref{ESVEA}.
Therefore, we opt to define the pulse's width in terms of the sech
approximation \eqref{Esech}. In this case, the ratio between the FWHM
duration, $T_{S}$, and the half width at the maximum value of the $\mathrm{%
sech}$ function, which is precisely $\tau_{S}$ in normalized form, is well
known to yield the $2\ln (2+\sqrt{3})$ factor.


Based on these considerations, we obtain the following duration of the
electromagnetic pulse:
\begin{equation}
T_{S}=2\ln (2+\sqrt{3})\frac{\sqrt{\varepsilon }}{\omega _{0}\sigma U}\sqrt{
\frac{1-U^{2}}{1-\Omega ^{2}}}.  \label{25}
\end{equation}%
To summarize, the pulse's shape of is fully characterized by the
dimensionless speed $U$ and frequency of internal oscillations $\Omega $,
which determine the carrier frequency $\omega _{c}$ and characteristic
duration $T_{S}$.

The number of cycles in the pulse can then be defined as $N_{p}=T_{S}/T_{c}$%
, where $T_{c}=2\pi /\omega _{c}$ is the period of the carrier wave. Then,
according to Eqs.~\eqref{24} and \eqref{25},
\begin{equation}
N_{p}=\frac{\ln \left( 2+\sqrt{3}\right) \Omega }{\pi \sqrt{1-\Omega ^{2}}U}.
\label{nbp}
\end{equation}%
It is thus seen that the parameter $U$ essentially defines the few- or
sub-cycle character of the pulse, from the slowly-varying-envelope
approximation at $U\rightarrow 0$ to the deeply sub-cycle configuration at $%
U\rightarrow 1$. Due to the above-mentioned discrepancy between $\omega _{c}$
and the actual maximum of the optical spectrum, $N_{p}$ defined as per Eq. (%
\ref{nbp}) underestimates the number of optical cycles, by approximately $%
8\% $ in the limit of the single-cycle regime ($N_{p}=1$), which is obtained
for $U=0.242$. At $U\rightarrow 1$, $N_{p}$ approaches $0.24$, while the
value computed from the maximum of the optical spectrum is $0.46$.
Obviously, the concept of the number of cycles is ambiguous in the sub-cycle
regime.

In the sub-cycle regime, with $T_{S}<T_{c}$, the oscillations of $\sin \chi $
cannot be considered as a carrier wave anymore. Then, the central frequency
of the pulse is mainly determined by the inverse $1/T_{S}$ of its duration,
as can be checked by numerically computing the optical spectrum, which
amounts to the computation of the Fourier transform of $E_{x}$. Hence, we
define the optical frequency as $\omega _{\mathrm{opt}}=\omega _{c}$ if $%
N_{p}>1$, i.e., if $U<0.242$, and $\omega _{\mathrm{opt}}=2\pi /T_{S}$
otherwise. The corresponding wavelength in vacuum is $\lambda _{\mathrm{opt}%
}=cT_{c}$ for $U<0.242$ and $\lambda _{\mathrm{opt}}=cT_{S}$ for $U>0.242$.
It can easily be checked that the maximum value of field $E_{x}$ is
\begin{equation}
E_{\mathrm{max}}=4E_{0}\frac{\sigma \sqrt{1-\Omega ^{2}}}{\sqrt{1-U^{2}}},
\label{eq:Em}
\end{equation}%
which increases with $U$, and diverges at $U\rightarrow 1$, hence, $U$ may
be also considered as a measure of the pulse's peak intensity, $I_{p}$.
Within the slowly-varying-envelope approximation,
\begin{equation}
I_{p}=\frac{c\sqrt{\epsilon }}{8\pi }E_{\mathrm{max}}^{2}=\frac{2c\sqrt{
\epsilon }E_{0}^{2}}{\pi }\frac{\sigma ^{2}(1-\Omega ^{2})}{1-U^{2}}.
\label{eq:Ip}
\end{equation}%
Although the actual intensity depends on the wave's velocity and may differ
from this expression, we will use Eq.~\eqref{eq:Ip} in the few- and subcycle
regimes.


\section{Transmission and reflection coefficients}

As a result of the interaction with the layer of increased electron
concentration, the initial electromagnetic pulse is generally split into
reflected and transmitted ones. The ratio of the energies of the transmitted
and reflected wave packets depends on various parameters, including the
initial characteristics of the incident pulse, and also parameters of the
scattering layer. To quantify this, we calculate the transmission and
reflection coefficients, $K_{\mathrm{pass}}$ and $K_{\mathrm{refl}}$, as~per
\cite{35}
\begin{equation}
K_{\mathrm{pass}}=\frac{\int_{0}^{+\infty }d\zeta \int_{-\infty }^{+\infty
}d\upsilon \int_{-\infty }^{+\infty }d\xi I(\xi ,\upsilon ,\zeta ,\tau
_{\infty })}{\int_{-\infty }^{+\infty }d\zeta \int_{-\infty }^{+\infty
}d\upsilon \int_{-\infty }^{+\infty }d\xi I(\xi ,\upsilon ,\zeta ,\tau
_{\infty })},  \label{26}
\end{equation}

\begin{equation}
K_{\mathrm{refl}}=\frac{\int_{-\infty }^{0}d\zeta \int_{-\infty }^{+\infty
}d\upsilon \int_{-\infty }^{+\infty }d\xi I(\xi ,\upsilon ,\zeta ,\tau
_{\infty })}{\int_{-\infty }^{+\infty }d\zeta \int_{-\infty }^{+\infty
}d\upsilon \int_{-\infty }^{+\infty }d\xi I(\xi ,\upsilon ,\zeta ,\tau
_{\infty })}.  \label{27}
\end{equation}%
In Eqs.~\eqref{26} and~\eqref{27}, $\tau _{\infty }$ corresponds to any time
after the establishment of the stable propagation of the electromagnetic
pulse, after its interaction with the layer, when the pulse is already at a
sufficiently large distance away from it, so that the field energy density
in this layer is negligible in comparison to the maximum energy density of
the field, i.e., condition $I(\xi ,\upsilon ,0,\tau _{\infty })\ll I_{%
\mathrm{max}}\equiv \text{max}\left\{ I(\xi ,\upsilon ,\zeta ,\tau _{\infty
})\right\} $ holds. Coefficient $K_{\mathrm{pass}}$, defined as per Eq.~%
\eqref{26}, may be interpreted as the ratio of the energy of the wave packet
passing the layer of increased electron concentration to the total field
energy in the volume of the sample. Similarly, $K_{\mathrm{refl}}$, defined
by Eq.~\eqref{27}, is the share of the energy of the reflected wave packet
in the sum of the energies of the transmitted and reflected packets.

The system considered in this paper is conservative since the collisionless
approximation is assumed. Therefore, the energy conservation law imposes
constraint
\begin{equation}
K_{\mathrm{refl}}+K_{\mathrm{pass}}=1.  \label{28}
\end{equation}%
Note that, when condition $I(\xi ,\upsilon ,0,\tau _{\infty })\ll I_{\mathrm{%
\ max}}$ is satisfied, and also by virtue of the energy conservation law,
coefficients $K_{\mathrm{pass}}$ and $K_{\mathrm{refl}}$ are
time-independent.

If the energy of the wave packet propagating in the original direction
significantly exceeds the energy of the wave packet reflected from the
inhomogeneity layer, i.e., $K_{\mathrm{pass}}\gg K_{\mathrm{refl}}$, we
assume that the pulse has passed through this layer. In the opposite case,
when the energy of the reflected wave packet significantly prevails over the
energy of the transmitted one, i.e., $K_{\mathrm{pass}}\ll K_{\mathrm{refl}}$%
, we categorize the pulse as reflected. For certain values of the system's
parameters---in particular, at some \textquotedblleft threshold" value of
the initial peak intensity, $I_{p_{\mathrm{thr}}}$, of the incident
pulse---it may split in two wave packets with approximately equal energies,
which propagate in opposite directions after the interaction with the
inhomogeneity layer, thereby corresponding to $K_{\mathrm{pass}}\approx K_{
\mathrm{refl}}$.


\section{Numerical analysis and discussion of the results}


\subsection{System parameters}

For modeling, we used the following realistic values of the parameters of
CNTs of the zig-zag type $(m,0)$: $m=7$, $\gamma _{0}=2.7$ eV, $b=1.42\times
10^{-8}$ cm, $d_{x}\approx 2.13\times 10^{-8}$ cm, $n_{\mathrm{bias}%
}=10^{18} $ cm$^{-3}$. We assume that the CNT array is embedded in a
dielectric matrix, and the resulting effective dielectric constant of the
system is $\varepsilon =4$, the calculation of coefficients $G_{r}$ given by
Eq.~\eqref{5} in the expression for the conduction current density~\eqref{4}
is performed for temperature $T=293$ K (see, for example, Refs.~\cite{30,35}%
).

We note that the use of the collisionless approximation (which makes it
possible to neglect dissipative effects, as mentioned above) is justified
when the evolution time interval is smaller than the relaxation time $t_{%
\mathrm{\ rel}}$. In the course of the respective time, $t_{\mathrm{rel}%
}\simeq 10^{-11}$ s, the electromagnetic pulse passes distance $z\leq ct_{%
\mathrm{rel }}/\sqrt{\varepsilon }\simeq 0.15$ cm.

When simulating the interaction of the pulse with a layer of the increased
electron concentration, we vary values of the parameters within a wide
range. In particular, parameter $\eta _{\mathrm{imp}}^{\mathrm{max}}=n_{
\mathrm{imp}}^{\mathrm{max}}/n_{\mathrm{bias}}$, corresponding to the
maximum concentration of electrons in the inhomogeneity layer, ranges from $%
2 $ to $100$. Next, the layer's dimensionless thickness, $\delta \zeta _{%
\mathrm{imp}}$, is varied between $0.05$ to $0.5$. Parameter $U$ of the
electromagnetic pulse approaching the layer is chosen in the interval of $%
U\in (0.5;0.999)$. We note that, at velocities $U<0.5$, in the course of
time $\sim t_{\mathrm{rel}}$, the pulse passes a negligible distance, which
is much smaller than its own spatial width along the $\zeta $-axis.
Velocities corresponding to $U>0.999$ are not considered here because of
limitations imposed by the numerical scheme.

Frequency $\Omega $ of the internal vibrations of the breather-like
electromagnetic pulse is varied in the interval of $\Omega \in (0.1;0.9)$.
As it decreases, the width of the pulse along the $\zeta $-axis decreases
too, although for $\Omega \leq 0.5$ the corresponding change of the pulse's
profile is insignificant. At $\Omega >0.7$, the width of the pulse becomes
comparable with dimensions of the numerical grid, which was chosen in
accordance with the characteristic size of real samples of CNT arrays.
Lastly, transverse width $w_{0}$ of the pulse is varied between $0.5$ and $%
2.0$, which leads to no qualitative difference in the character of the
interaction of the electromagnetic pulse with the region of increased
electron concentration.

\subsection{Scenarios for the interaction of the electromagnetic pulse with
a layer of increased electron concentration}


Equations~\eqref{7},~\eqref{10}, and~\eqref{12} do not admit analytical
solutions, therefore we carried out numerical simulations to study the
propagation of the electromagnetic pulse in the CNT array. To solve this
system of equations with initial conditions~\eqref{15},~\eqref{16}, and~%
\eqref{17}, we used an explicit finite-difference three-layer scheme of the
\textquotedblleft cross" type described in Refs.~\cite{48,49,50}, which was
adapted by us for the 3D model, using the approach developed in Ref.~\cite%
{35}. Here, we do not describe it in detail, as the numerical scheme and the
computational algorithm are similar to those presented in a detailed form in
Ref.~\cite{35} for the 2D geometry. As a result of the computations, we\
have found fields $\Psi (\xi ,\upsilon ,\zeta ,\tau )$, $\eta (\xi ,\upsilon
,\zeta ,\tau )$, and $\Phi (\xi ,\upsilon ,\zeta ,\tau )$, and also
calculated the distribution of the field energy density at each instant of
time, using Eq.~\eqref{14}.

The simulations reveal that, depending on values of certain system's
parameters, different scenarios of the interaction of the ultrashort pulse
with the layer of increased electron concentration are possible. The pulse
may either pass the layer or bounce back from it. Control parameters, whose
values determine the result of the interaction, are characteristics of the
electromagnetic pulse (in particular, the speed at which it is approaching
the inhomogeneity layer) and of the layer itself (its thickness and the
concentration of conduction electrons in it). Passing the layer is
facilitated both by the increase in the peak intensity of the incident
pulse, and by the decrease in the thickness of the layer and concentration
of electrons in it. In fact, similar scenarios of the interaction of the
pulse with the layer are produced by varying all control parameters.

\begin{figure}[tbp]
\includegraphics[width=1.0\textwidth]{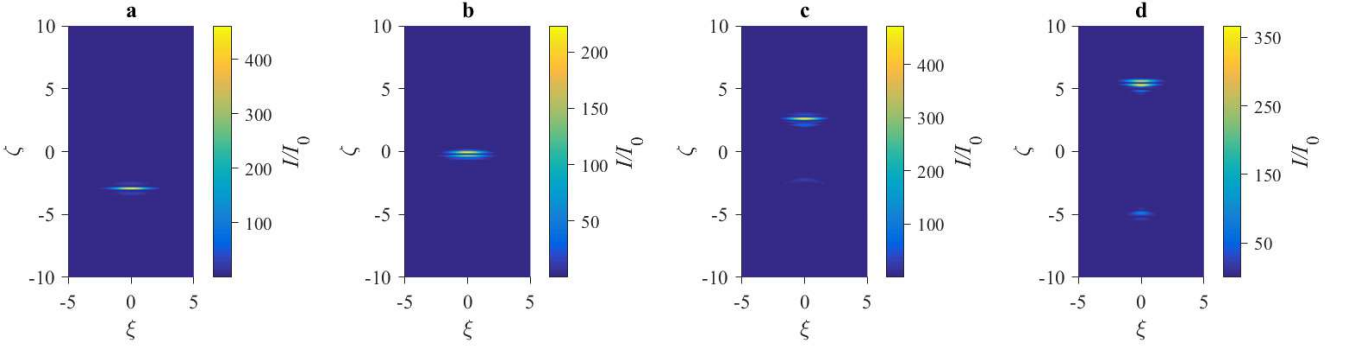}
\caption{Distribution of the energy density of the electric field $I(\protect%
\xi ,\protect\upsilon _{0},\protect\zeta ,\protect\tau )$ in the array of
nanotubes at various moments of dimensionless time $\protect\tau =\protect%
\omega _{0}t/\protect\sqrt{\protect\varepsilon }$, in the course of the
passage of the laser pulse through the layer of high electron concentration,
located at $\protect\zeta =0$: (a) $\protect\tau =0$, (b) $\protect\tau =3.0$
, (c) $\protect\tau =6.0$, (d) $\protect\tau =9.0$. Dimensional coordinates $%
\protect\xi =x\protect\omega _{0}/c$ and $\protect\zeta =z\protect\omega %
_{0}/c$ are plotted along the horizontal and vertical axes. Values of $%
I/I_{0}$ are mapped with the help of the color scale, yellow and blue areas
corresponding, respectively, to the maximum and minimum values of the energy
density. }
\label{fig3}
\end{figure}

In what follows, we present a number of key results for the propagation of
the ultrashort pulse in the inhomogeneous CNT\ array for different values of
parameters of the pulse and inhomogeneity layer. Figures~\ref{fig3} and~\ref%
{fig4} display the results for the ultrashort pulse interacting with the
layer at different values of the pulse's parameter $U$, while other
parameters remain constant: the dimensionless frequency of internal
oscillations is $\Omega =0.5$, transverse pulse's width $w_{0}=1.75$, and
characteristics of the inhomogeneity layer $\eta _{\mathrm{imp}}^{ \mathrm{%
max}}=30$, $\delta \zeta _{\mathrm{imp}}=0.1$.

Figure~\ref{fig3} illustrates the passage of the electromagnetic pulse
through the layer of increased electron concentration. The temporal
narrowness of the incident pulse and its peak intensity are determined by $%
U=0.99$. The duration of such a pulse is $T_{S}\simeq 12.8\times 10^{-15}$~s
(see Eq.~\eqref{25}), which corresponds to the optical frequency $\omega _{%
\mathrm{opt}}\simeq 4.9\times 10^{14}$~s$^{-1}$, and wavelength $\lambda _{%
\mathrm{opt}}\simeq 2.8$~$\mu $m, at the lower limit of the mid-infrared
band. The maximum amplitude of the electric field of the pulse is $E_{%
\mathrm{max}}\simeq 8.6\times 10^{4}$ statvolt/cm $\simeq 2.6\times 10^{9}$%
~V/m, which follows from Eq.~\eqref{eq:Em}. This electric-field amplitude
corresponds to a peak intensity $I_{p}\simeq 1.76\times 10^{19}$~erg$\cdot $%
cm$^{2} \cdot $s$^{-1}\simeq 1.76$ TW/cm$^{2}$.


Figure~\ref{fig3} shows the distribution of the energy density of the
electric field $I(\xi ,\upsilon _{0},\zeta ,\tau )$ (see Eq.~\eqref{14}) in
the cross-section $\left( \xi ,\zeta \right) $ (at $\upsilon =\upsilon _{0}$%
), at different values of the dimensionless time, $\tau =\omega _{0}t/\sqrt{
\varepsilon }$. The energy density of the electric field is represented by
ratio $I/I_{0}$, using a suitable colormap, with blue and yellow areas
corresponding, respectively, to minimum and maximum values. For values of
the system's parameters selected above, the unit along the $\xi $- and $%
\zeta $-axes corresponds to distance $\simeq 4.2\times 10^{-4}$ cm. Note
that we show the distribution of $I/I_{0}$ only in the cross-section $\left(
\xi ,\zeta \right) $ (for $\upsilon =\upsilon _{0}$), as the pattern of the
distribution of the energy density in the cross-section $\left( \upsilon
,\zeta \right) $ is quite similar.

It can be seen from Fig.~\ref{fig3} that the electromagnetic pulse after
interacting with the inhomogeneity layer passes it and continues to stably
propagate in the medium in the original direction. Note that in this case,
only a negligible fraction of the initial electromagnetic pulse is
reflected, in the form of a wave packet with a small amplitude propagating
in the opposite direction. The transmission and reflection coefficients in
this case are $K_{\mathrm{pass}}\approx 0.8464$ and $K_{\mathrm{refl}%
}\approx 0.1536$, respectively, satisfying relation $K_{\mathrm{pass}}\gg
K_{ \mathrm{refl}}$, which allows us to speak mainly about the passage of
the layer of high electron concentration by the pulse.

\begin{figure}[tbp]
\includegraphics[width=1.0\textwidth]{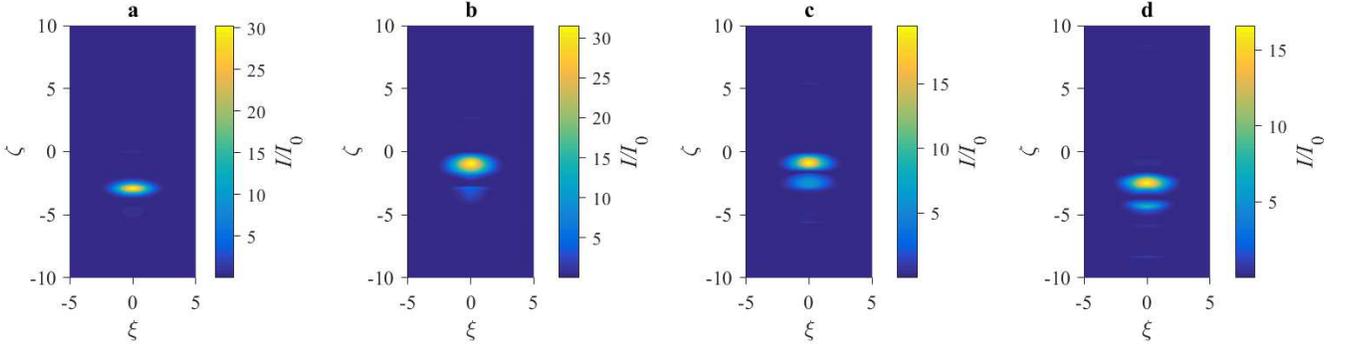}
\caption{The distribution of the energy density of the electric field, $I(
\protect\xi ,\protect\upsilon _{0},\protect\zeta ,\protect\tau )$ in the
array of nanotubes at different values of dimensionless time $\protect\tau $
, when the laser pulse is reflected from the layer of increased electron
concentration placed at $\protect\zeta =0$: (a) $\protect\tau =0$, (b) $%
\protect\tau =3.0$, (c) $\protect\tau =6.0$, (d) $\protect\tau =9.0$. The
notation is the same as in Fig.~\protect\ref{fig3}.}
\label{fig4}
\end{figure}

Figure~\ref{fig4} shows the reverse situation, namely the reflection of the
electromagnetic pulse, with $U=0.80$, from the layer of increased electron
concentration. These values of the parameters of the electromagnetic pulse
correspond to duration $T_{S}\simeq 6.7\times 10^{-14}$~s, optical frequency
$\omega _{\mathrm{opt}}\simeq 9.3\times 10^{13}$~s$^{-1}$, and wavelength $%
\lambda _{\mathrm{opt}}\simeq 20~\mu $m, in the mid-infrared range, and the
maximum of the electric field $E_{\mathrm{max}}\simeq 2.0\times 10^{4}$%
~statvolt/cm$\simeq 6.1\times 10^{8}$ V/m, which corresponds to peak
intensity $I_{p}\simeq 9.75\times 10^{17}$~erg$\cdot $cm$^{2}\cdot $s$%
^{-1}\simeq 97.5$~GW/cm$^{2}$. Similarly to Fig.~\ref{fig3}, this figure
shows the distribution of the energy density of the electric field $I(\xi
,\upsilon _{0},\zeta ,\tau )$ in cross-section $\left( \xi ,\zeta \right) $
of the CNT array at various values of dimensionless time $\tau $.
The electromagnetic pulse, having an insufficiently high initial peak
intensity, does not pass this region, bouncing back and propagating in the
reverse direction. The respective transmission and reflection coefficients
are $K_{\mathrm{pass}}\approx 0.075$ and $K_{\mathrm{refl}}\approx 0.9925$.
Relation $K_{\mathrm{pass}}\ll K_{\mathrm{refl}}$ in this case may be
interpreted as satisfying the criterion for the reflection of the laser
pulse from the layer of an increased electron concentration.

As shown by the numerical analysis, the possibility of the passage of the
electromagnetic pulse through the layer of the high electron concentration
depends not only on the peak intensity of the incident pulse, but also on
parameters of the layer, such as its thickness, $\delta \zeta _{\mathrm{imp}%
} $, and the maximum reduced concentration of electrons in it, $\eta _{%
\mathrm{\ imp}}^{\mathrm{max}}$.

\begin{figure}[tbp]
\includegraphics[width=1.0\textwidth]{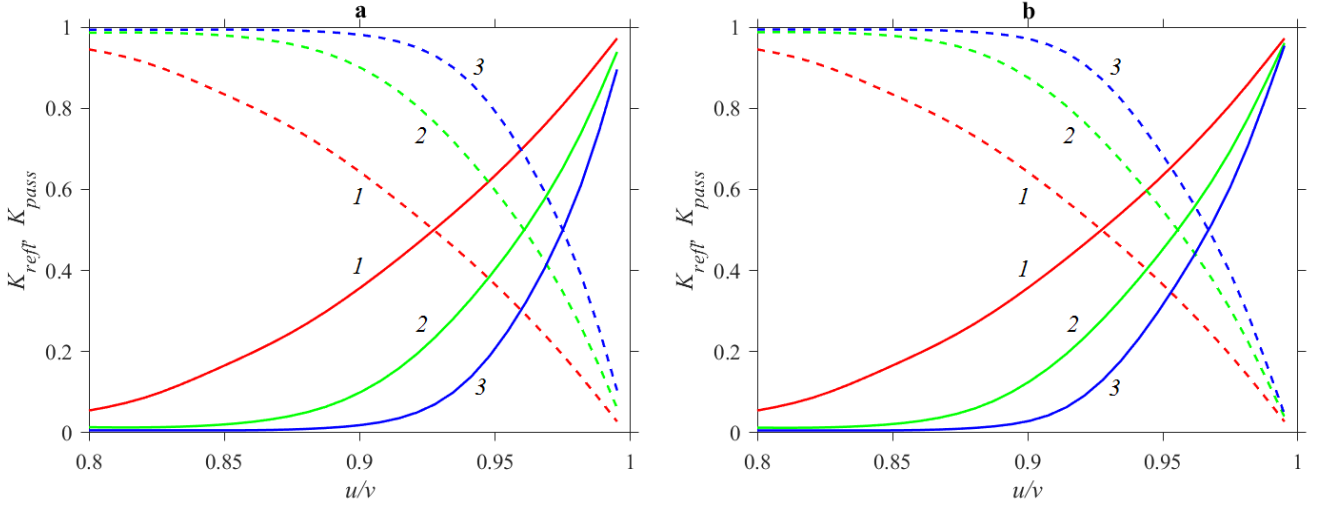}
\caption{Transmission and reflection coefficients, $K_{\mathrm{pass}}$ and $%
K_{\mathrm{refl}}$ (solid and dashed curves, respectively) vs. parameter $U$
, which characterizes the narrowness and intensity strength of the incident
electromagnetic pulse when it interacts with the layer of increased electron
concentration: (a) for increasing concentration, namely, for $\protect\delta
\protect\zeta _{\mathrm{imp}}=0.1$: {\emph 1} (red) -- $\protect\eta _{\mathrm{imp}}^{
\mathrm{max}}=20$, {\emph 2} (green) -- $\protect\eta _{\mathrm{imp}}^{\mathrm{max}}=30$
, {\emph 3} (blue) -- $\protect\eta _{\mathrm{imp}}^{\mathrm{max}}=40$; and (b) for
increasing values of the layer's thickness, namely, for $\protect\eta _{
\mathrm{imp}}^{\mathrm{max}}=20$: {\emph 1} (red) -- $\protect\delta \protect\zeta _{
\mathrm{imp}}=0.10$, {\emph 2} (green) -- $\protect\delta \protect\zeta _{\mathrm{imp}
}=0.15$, {\emph 3} (blue) -- $\protect\delta \protect\zeta _{\mathrm{imp}}=0.20$.}
\label{fig5}
\end{figure}

Figure~\ref{fig5} shows the dependence of the reflection and transmission
coefficients of the electromagnetic pulse on parameter $U$ that determines
its initial peak intensity, longitudinal width and duration. As $U$
increases, the transmission coefficient $K_{\mathrm{pass}}$ (the solid
curve) increases too and, accordingly, the reflection coefficient $K_{%
\mathrm{refl}}$ (the dotted curve) decreases. The value $U=U_{\mathrm{thr}}$
of $U$ is such that the two curves intersect, thereby implying the equality
between the transmission and reflection coefficients:
\begin{equation}
K_{\mathrm{pass}}(U_{\mathrm{thr}})=K_{\mathrm{refl}}(U_{\mathrm{thr}}).
\end{equation}%
The latter equation fully defines the threshold value $U_{\mathrm{thr}}$.
For values of parameters used in Figs.~\ref{fig3} and~\ref{fig4}, namely $%
\eta _{\mathrm{imp}}^{\mathrm{max}}=30$ and $\delta \zeta _{\mathrm{imp}%
}=0.1 $, the threshold is $U_{\mathrm{thr}}\simeq 0.96$, which corresponds
to the peak intensity $I_{p_{\mathrm{thr}}}\simeq 4.6\times 10^{18}$~erg$%
\cdot $cm$^{2}\cdot $s$^{-1}=460$~GW/cm$^{2}$. At $U\simeq U_{\mathrm{thr}}$%
, the electromagnetic pulse can be divided in two approximately identical
wave packets, one of which continues to move in the original direction,
while the other bounces back from the layer of increased electron
concentration. Calculations show that the transmission coefficient $K_{%
\mathrm{pass}}$ increases, and the reflection coefficient $K_{\mathrm{refl}}$
decreases with the increase of $U$, while other parameters are kept
constant. Further, $K_{\mathrm{pass}}$ decreases, and $K_{\mathrm{refl}}$
increases with the increase of any parameter of the inhomogeneity layer,
\textit{viz}., $\eta _{\mathrm{imp}}^{ \mathrm{max}}$ [see Fig.~5(a)] and $%
\delta \zeta _{\mathrm{imp}}$ [see Fig.~5(b)], at a fixed value of $U$.

Thus, the result of the interaction of the laser pulse with the layer of the
increased electron concentration depends on various parameters, including
the initial pulse's parameter $U$, which controls peak intensity $I_{p}$,
optical frequency $\omega _{\mathrm{opt}}$, duration $T_{S}$, the maximum
reduced electron concentration in the inhomogeneity layer, $\eta _{\mathrm{\
imp}}^{\mathrm{max}}$, and the characteristic thickness of this layer, $%
\delta \zeta _{\mathrm{imp}}$.

Qualitatively, these dependencies have a simple physical explanation. As can
be seen from Eq.~\eqref{4}, the current density induced by the pulse is
proportional to the conduction electrons concentration. Therefore, with the
increase of the carrier concentration of conduction electrons. Therefore,
with the increase of the carrier concentration in the impurity band, the
current that creates the pulse also increases, i.e., the impurity region
becomes more conductive. Even at the level of linearized equations, using
the Fresnel formulas for the reflection and transmission coefficients, it is
clear that a more conducting medium reflects the electromagnetic wave more
efficiently. The obtained dependencies on the speed of the pulses may be
qualitatively understood too. As the peak intensity increases, the velocity
also increases, reducing the time of the interaction with the impurity band,
hence the time available for generating the pulse running in the reverse
direction is reduced. Indeed, as can be seen from Eq.~\eqref{eq:Ip}, the
higher the intensity, the higher the pulse velocity is. From Eq.~%
\eqref{Esech}, one can notice that a higher pulse velocity leads to a
reduction in the longitudinal width of the pulse, which consequently yields
a shorter pulse duration [Eq.~\eqref{25}]. In summary, a decrease in the
pulse duration and longitudinal width results in a shorter time interval
during which the pulse interacts and passes through the region with an
increased electron concentration.

Another explanation for the particular pulse dynamics observed in the
presence of the layer of inhomogeneity is that a higher peak intensity of
the pulse makes it easier to overcome the effective potential barrier
induced by the impurity band. In Refs.~\cite{50a,50b} an energy-based
analysis of the interaction between the pulse and a layer of increased
electron concentration in a semiconductor is reported. The energy of the
soliton, necessary to overcome the effective potential barrier created by
the impurity region, was defined through the effective Lagrangian of the
soliton in a vicinity of an impurity, in the framework of a model based on
an inhomogeneous (perturbed) wave equation of the sine-Gordon type, similar
to the governing equation used in the present work. In Refs.~\cite{50a,50b}
it was established that larger values of the pulse's velocity---or,
equivalently, of the peak intensity---and smaller values of the pulse
duration favor the passage of a pulse through the region of increased
conductivity.

In connection to these results, it is relevant to compare them with known
results obtained for the 1D sine-Gordon equation with impurities~\cite%
{51,52,53,54}. Conclusions formulated in those works indicate that,
colliding with an inhomogeneity, the breather is either attenuated, due to
the emission of linear waves, or splits into a kink-antikink pair (note that
an essential part of those above results was obtained analytically). The
difference demonstrated by our results is related to two central factors:
(i) the three-dimensionality of our problem, which modifies the dispersion
law, and (ii) the nonlinearity of our medium, represented by multiple sine
terms. Because of the nonintegrability of the present model, our
quasi-soliton suffers radiation losses, which, however, become significant
for times much larger than those we are considering here, namely, at times
when the relaxation effects in the electronic subsystem become significant.
The effects of the three-dimensionality, in particular, imply the necessity
to redefine the topological charge, which plays a major role in the dynamics
of the 1D models. In the 1D case, the topological charge acts as a
\textquotedblleft selection rule", which prohibits certain decay mechanisms.
In the present case, these rules do not apply, because of the
three-dimensionality. While the breather-like solutions suffer some
radiation losses, as mentioned above, it is not seen in the
reflection/transmission coefficients, as the integration is carried out
throughout the entire spatial domain, taking the contribution from the
radiation field into account.

Thus, based on the results produced by our numerical analysis, it can be
asserted that the electromagnetic pulses with relatively low peak
intensities cannot pass the layer of increased electron concentration, while
the pulses with peak intensities significantly exceeding a certain threshold
value overcome the repelling layer. As the peak intensity of the pulse
increases, the possibility of its passage through the layer increases too.
In other words, an increase in the optical frequency and a decrease in the
duration of the electromagnetic pulse contribute to the ability of pulse to
pass the layer.

The threshold value of parameter $U_{\mathrm{thr}}$ of the electromagnetic
pulse, in turn, depends on a number of factors, including parameters $\eta
_{ \mathrm{imp}}^{\mathrm{max}}$ and $\delta \zeta _{\mathrm{imp}}$ of the
layer of increased electron concentration. The possibility of the passage of
the inhomogeneity layer by the pulse increases with a decrease in these
parameters. These findings refine and generalize conclusions obtained in
previous works devoted to the study of the interaction of extremely short
pulses with layers of an increased electron concentration in CNT arrays,
which were based on 2D models \cite{34,35}. The system considered in the
present work acts as a \textquotedblleft filter" for extremely short
electromagnetic pulses, selectively transmitting narrow ones (with higher
optical frequencies), and reflecting pulses of longer durations, with lower
frequencies. This effect may be used as a basis for the operation of optical
logic elements and laser field control devices, as well as in the technology
of nondestructive quality control of electronic elements based on CNTs.


\section{Conclusions}


Key results of this work are summarized as follows:

\begin{itemize}
\item[i)] It has been established that, as a result of the scattering of the
electromagnetic pulse on the layer of increased electron concentration in
the array of CNTs (carbon nanotubes), both the passage of the pulse through
the layer and reflection from it take place.

\item[ii)] The result of the interaction of the electromagnetic pulse with
the layer of increased electron concentration depends on values of the
system's parameters, including the speed (determined by the optical
frequency and duration) of the pulse, and also on characteristics of the
inhomogeneity layer (its thickness and the excess of the conduction electron
concentration with respect to the bulk array).

\item[iii)] The increase in the peak intensity (or increase in the optical
frequency, or decrease in the duration) of the electromagnetic pulse, as
well as the decrease in the thickness of the inhomogeneity layer and
concentration of conduction electrons in it, facilitate the passage of the
pulse through this layer.

\item[iv)] After interacting with the layer of high electron concentration,
the electromagnetic pulse retains its characteristics, remaining an
oscillating bipolar wave packet that can propagate steadily over distances
that are noticeably larger than its dimensions along the direction of motion.
\end{itemize}

\begin{acknowledgments}
A. V. Zhukov and R. Bouffanais are financially supported by the SUTD-MIT
International Design Centre (IDC). N. N. Rosanov acknowledges the support
from the Russian Foundation for Basic Research, Grant 16-02-00762, and from
the Foundation for the Support of Leading Universities of the Russian
Federation (Grant 074-U01). M. B. Belonenko acknowledges support from the
Russian Foundation for Fundamental Research. E. G. Fedorov is grateful to
Prof. Tom Shemesh for his generous support. B. A. Malomed appreciates
hospitality of the School of Electrical and Electronic Engineering at the
Nanyang Technological University (Singapore).

All the authors contributed equally to this work.
\end{acknowledgments}


\begin{thebibliography}{99}
\bibitem{1} R. H. Baughman, A. A. Zakhidov, and W. A. de Heer, Science
\textbf{\ 297}, 787 (2002).

\bibitem{2} S. Iijima, Nature \textbf{354}, 56 (1991).

\bibitem{3} S. Iijima and T. Ichihashi, Nature \textbf{363}, 603 (1993).

\bibitem{4} M. S. Dresselhaus, G. Dresselhaus, P. Eklund, \textit{The
science of fullerenes and carbon nanotubes} (Elsevier, 1996).

\bibitem{5} \textit{Carbon nanotubes, preparation and properties}, T. W.
Ebbesen, Ed. (CRC Press, 1996).

\bibitem{6} R. Saito, G. Dresselhaus, and M. S. Dresselhaus, \textit{\
Physical properties of carbon nanotubes} (World Scientific, 1998).

\bibitem{7} P. J. F. Harris, \textit{Carbon Nanotubes and Related
Structures: New Materials for the Twenty-First Century} (Cambridge
University Press, 1999).

\bibitem{8} S. A. Maksimenko and G. Ya. Slepyan, J. Comm. Techn. Elect.
\textbf{\ 47}, 261 (2002).

\bibitem{9} S. A. Maksimenko and G. Ya. Slepyan, in \textit{Handbook of
Nanotechnology. Nanometer Structure: Theory, Modeling, and Simulation} (SPIE
Press, Bellingham, 2004).

\bibitem{10} A. V. Zhukov, R. Bouffanais, M. B. Belonenko, and E. G.
Fedorov, Mod. Phys. Lett. B \textbf{27}, 1350045 (2013).

\bibitem{11} M. B. Belonenko and E. G. Fedorov, Russ. Phys. J. \textbf{55},
83 (2012).

\bibitem{12} M. B. Belonenko, E. V. Demushkina, and N. G. Lebedev, J. Russ.
Laser Res. \textbf{27}, 457 (2006)

\bibitem{13} S. A. Akhmanov, V. A. Vysloukh, and A. S. Chirkin, \textit{\
Optics of femtosecond laser pulses} (American Institute of Physics, 1992).

\bibitem{14} A. M. Zheltikov, Phys. Usp. \textbf{50}, 705 (2007).

\bibitem{15} S. V. Sazonov, Bull. Russ. Acad. Sci.: Physics \textbf{75}, 157
(2011).

\bibitem{16} H. Leblond and D. Mihalache, Phys. Rep. \textbf{523}, 61 (2013).

\bibitem{17} G. Mourou, S. Mironov, E. Khazanov, and A. Sergeev, Eur. Phys.
J. Special Topics, \textbf{223}, 1181 (2014).

\bibitem{18} M. Kolesik and J. V. Moloney, Rep. Prog. Phys. \textbf{77},
016401 (2014).

\bibitem{19} D. J. Frantzeskakis, H. Leblond, and D. Mihalache, Rom. J.
Phys. \textbf{59}, 767 (2014).

\bibitem{20} B. A. Malomed, D. Mihalache, F. Wise, and L. Torner, J. Opt.
B.: Quantum Semiclass. Opt. \textbf{7}, R53 (2005).

\bibitem{21} D. Mihalache, Rom. J. Phys. \textbf{57}, 352 (2012).

\bibitem{22} D. Mihalache, Rom. J. Phys. \textbf{59}, 295 (2014).

\bibitem{23} M. B. Belonenko, N. N. Yanushkina, and E. G. Fedorov, Bull.
Russ. Acad. Sci.: Physics \textbf{76}, 1326 (2012).

\bibitem{23a} B. Malomed, L. Torner, F. Wise, and D. Mihalache, J. Phys. B:
At. Mol. Opt. Phys. \textbf{49}, 170502 (2016).

\bibitem{23b} D. Mihalache, Rom. Rep. Phys. \textbf{69}, 403 (2017).

\bibitem{24} M. B. Belonenko, N. G. Lebedev, and E. N. Nelidina, Phys. Wave
Phen. \textbf{19}, 39 (2011).

\bibitem{25} E. G. Fedorov, A. V. Zhukov, M. B. Belonenko, and T. F. George,
Eur. Phys. J. D \textbf{66}, 219 (2012).

\bibitem{26} H. Leblond and D. Mihalache, Phys. Rev. A \textbf{86}, 043832
(2012).

\bibitem{27} E. G. Fedorov, A. V. Pak, and M, B. Belonenko, Phys. Sol. State
\textbf{56}, 2112 (2014).

\bibitem{28} M. B. Belonenko and E. G. Fedorov, Phys. Sol. State \textbf{55}
, 1333 (2013).

\bibitem{29} A. V. Zhukov, R. Bouffanais, E. G. Fedorov, and M. B.
Belonenko, J. Appl. Phys. \textbf{114}, 143106 (2013).

\bibitem{30} A. V. Zhukov, R. Bouffanais, B. A. Malomed, H. Leblond, D.
Mihalache, E. G. Fedorov, N. N. Rosanov, and M. B. Belonenko, Phys. Rev. A
\textbf{94}, 053823 (2016).

\bibitem{31} E. G. Fedorov, N. N. Konobeeva, and M. B. Belonenko, Russ. J.
Phys. Chem. B, \textbf{8}, 409 (2014).

\bibitem{32} M. B. Belonenko, N. G. Lebedev, and A. S. Popov, JETP Lett.
\textbf{91}, 461 (2010).

\bibitem{33} M. B. Belonenko, A. S. Popov, and N. G. Lebedev, Tech. Phys.
Lett. \textbf{37}, 119 (2011).

\bibitem{34} A. V. Zhukov, R. Bouffanais, E. G. Fedorov, and M. B.
Belonenko, J. App. Phys. \textbf{115}, 203109 (2014).

\bibitem{35} A. V. Zhukov, R. Bouffanais, H. Leblond, D. Mihalache, E. G.
Fedorov, and M. B. Belonenko, Eur. Phys. J. D \textbf{69}, 242 (2015).

\bibitem{35a} S. V. Kryuchkov, E. V. Kaplya, J. Commun. Technol. Electron
\textbf{51}, 359 (2006).

\bibitem{36} M. B. Belonenko, S. Yu. Glazov, and N. E. Meshcheryakova, Opt.
Spectr. \textbf{108}, 774 (2010).

\bibitem{Lagendijk} B. A. van Tiggelen, A. Tip and A. Lagendijk, J. Phys. A:
Math. Gen. \textbf{26}, 1731 (1993).

\bibitem{37} L. D. Landau, E. M. Lifshitz, and L. P. Pitaevskii, \textit{\
Electrodynamics of Continuous Media}, 2nd Ed. (Elsevier, Oxford, 2004).

\bibitem{38} L. D. Landau and E. M. Lifshitz, \textit{The Classical Theory
of Fields}, 4th Ed. (Butterworth-Heinemann, Oxford, 2000).

\bibitem{39} E. M. Epshtein, Fiz. Tverd. Tela \textbf{19}, 3456 (1976).

\bibitem{40} E. M. Epshtein, Fiz. Tech. Polupr. \textbf{14}, 2422 (1980)
[Sov. Phys. Semiconductors 14, 1438 (1980)].

\bibitem{41} G. A. Korn and T. M. Korn, \textit{Mathematical Handbook for
Scientists and Engineers} (McGraw Hill, New York, 1968).

\bibitem{42} Yu. S. Kivshar and B. A. Malomed, Rev. Mod. Phys. \textbf{61},
763 (1989).

\bibitem{43} \textit{Femtosecond Laser Pulses: Principles and Experiments},
Claude Rulli{\'e}re, ed. (Springer-Verlag, Berlin, 1998).

\bibitem{44} A. N. Pikhtin, \textit{Optical and Quantum Electronics} (High
School Publishers, Moscow, 2001).

\bibitem{45} A.M. Goncharenko, \textit{Gaussian light beams} (Nauka i
Tekhnika, Minsk, 1976).

\bibitem{46} M. B. Belonenko and E. G. Fedorov, Russ. Phys. J. \textbf{55},
436 (2012).

\bibitem{47} A. V. Zhukov, R. Bouffanais, M. B. Belonenko, and E. G.
Fedorov, Mod. Phys. Lett. B \textbf{27}, 1350045 (2013).

\bibitem{48} N. N. Kalinkin, \textit{Numerical Methods} (Nauka, Moscow,
1978).

\bibitem{49} S. E. Koonin, \textit{Computational Physics : Fortran Version}
( Ingram Publisher Services, Boulder, United States, 1998).

\bibitem{50} J. W. Thomas, \textit{Numerical Partial Differential Equations
-- Finite Difference Methods} (Springer-Verlag, New York, 1995).

\bibitem{50a} S. V. Kryuchkov and E. G. Fyodorov, Laser Phys. \textbf{12},
no. 7, 1037 (2002).

\bibitem{50b} S. V. Kryuchkov and E. G. Fyodorov, Optics and Spectroscopy
\textbf{94}, no. 2, 225 (2003).

\bibitem{51} B. A. Malomed, Physica D \textbf{27}, 113 (1987).

\bibitem{52} B. A. Malomed, Phys. Lett. A \textbf{120}, 28 (1987).

\bibitem{53} V. E. Zakharov, S. V. Manakov, S. P. Novikov, and L. P.
Pitaevskii, \textit{Theory of Solitons} (Consultants Bureau, New York,1984).

\bibitem{54} Yu. S. Kivshar and B. A. Malomed, Phys. Lett. A \textbf{115},
381 (1986).
\end{thebibliography}
\end{document}